\documentclass[twocolumn,aps,prb,showpacs,superscriptaddress]{revtex4-1}
\usepackage{graphics}
\usepackage{epsfig}
\usepackage{times}
\usepackage{bm}
\usepackage{color}

\usepackage{amsmath}	
\begin{document}

\title{Anderson localization versus charge-density-wave formation
in disordered electron systems}

\author{S.\ Nishimoto}
\affiliation{
Institute for Theoretical Solid State Physics, IFW Dresden,
01171 Dresden, Germany}

\author{S.\ Ejima}
\author{H.\ Fehske}
\affiliation{Institute of Physics,
             Ernst Moritz Arndt University Greifswald,
             17489 Greifswald, Germany}

\date{\today}

\begin{abstract}
We study the interplay of disorder and interaction effects 
including bosonic degrees of freedom in the framework 
of a generic one-dimensional transport model,
the Anderson-Edwards model. 
Using the density-matrix-renormalization group
technique, we extract the localization length and 
the renormalization of the Tomonaga-Luttinger-liquid 
parameter from the charge-structure factor by a elaborate 
sample-average finite-size scaling procedure.
The properties of the Anderson localized state 
can be described in terms of scaling relations 
of the metallic phase without disorder. We analyze  
how disorder competes with the charge-density-wave 
correlations triggered by the bosons 
and give evidence that disorder will destroy
the long-range charge-ordered state.
\end{abstract}
\maketitle
\section{Introduction}
Disorder is an inherent part of any solid state system.~\cite{Ga05} 
Low-dimensional materials are exceedingly susceptible to disorder. 
In one dimension (1D), theory predicts that all carriers 
are strongly localized for arbitrary energies and arbitrarily 
weak disorder. This holds for Anderson's non-interacting 
tight-binding Hamiltonian with a diagonal (i.e., on-site) 
random potential.~\cite{An58,FMSS85} The coherent backscattering 
from the randomly distributed impurities thereby transforms the 
metal into an insulator. 

In 1D the mutual interaction of the particles is likewise of significance; 
here even weak interactions can cause strong correlations. The instantaneous
Coulomb repulsion between the electrons, for instance,  
tends to immobilize the charge carriers as well. As a consequence, 
at half-filling, a Mott insulating (spin-density-wave) phase is 
energetically favored over the metallic state.~\cite{Mo90} 
The retarded electron-phonon coupling, on the other hand, may lead 
to structural distortions accompanied by polaron formation,~\cite{La33} 
and is the driving force behind the metal-to-Peierls transition, 
establishing a charge-density-wave (CDW) order.~\cite{Pe55}

An understanding of how disorder and interaction act together is 
of vital importance not only to discuss the metal-insulator 
itself but also to analyze the electronic properties 
of many quasi-1D materials of current interest, such as conjugated polymers, 
organic charge transfer salts, ferroelectric perovskites, 
halogen-bridged transition metal complexes, 
TMT[SF,TF] chains, Qn(TCNQ)$_2$ compounds, or, e.g., 
the quite recently studied vanadium dioxide Peierls-Mott 
insulator.~\cite{TNYS91,BS93,JS82,WOHPKL12} Carbon nanotubes~\cite{MAPHQTTCRNJ08}
and organic semiconductors~\cite{CF11} are other examples
where disorder and bosonic degrees of freedom are of importance.  
Regarding interacting bosons, ultracold atoms 
trapped in optical lattices offer the unique possibility to tune 
both the disorder and interaction strength.~\cite{RPLG12}  

Unfortunately the subtle interplay of disorder and interaction effects 
is one of the most challenging problems in solid state theory 
and---despite 50 years of intense research---is still an 
area of uncertainty, see Ref.~\onlinecite{Ab10} 
and references therein. In the 
limit of vanishing charge carrier density only the interaction 
with the lattice vibrations matters. Then Anderson disorder  
may affect the polaron self-trapping in a highly nontrivial way.~\cite{An72} 
This has been demonstrated for the Anderson-Holstein model within 
the statistical dynamical mean field and momentum average 
approximations.~\cite{BF02,EB12} 
At finite carrier density, the Mott-Anderson transition for Coulomb
correlated electrons was investigated by self-consistent mean-field 
theory in $\rm D=\infty$ and $\rm D=3$,~\cite{TL93,DK97,BHV05} as well as by 
variational Gutzwiller ansatz based approaches.~\cite{FGA11}  Electron-electron interactions may screen the disorder potential
in strongly correlated systems, stabilizing thereby metallicity.~\cite{TDAK03} 
Exact results are rare however.
In 2D, Lanczos and quantum Monte Carlo data suggest 
a disorder-induced stabilization of the pseudogap, also away from 
half-filling.~\cite{CCPS08} The density matrix renormalization group
(DMRG)~\cite{Wh92} allows the numerical exact calculation of 
ground-state properties of disordered, interacting fermion systems 
in 1D, on fairly large systems. 
Exploiting this technique, 
the properties of disordered Luttinger liquids have been analyzed
in the framework of the spinless fermion Anderson-$t$-$V$ 
model (A$tV$M)~\cite{SSSSE98} and the spinful Anderson-Hubbard 
model.~\cite{NS10}
    
In this paper, we address how many-body Anderson 
localization competes with CDW formation triggered by bosonic degrees of freedom 
in the framework of the Anderson-Edwards model (AEM). 

\section{Model}
\begin{figure}[t]
\centering
\includegraphics[clip,width=.8\columnwidth]{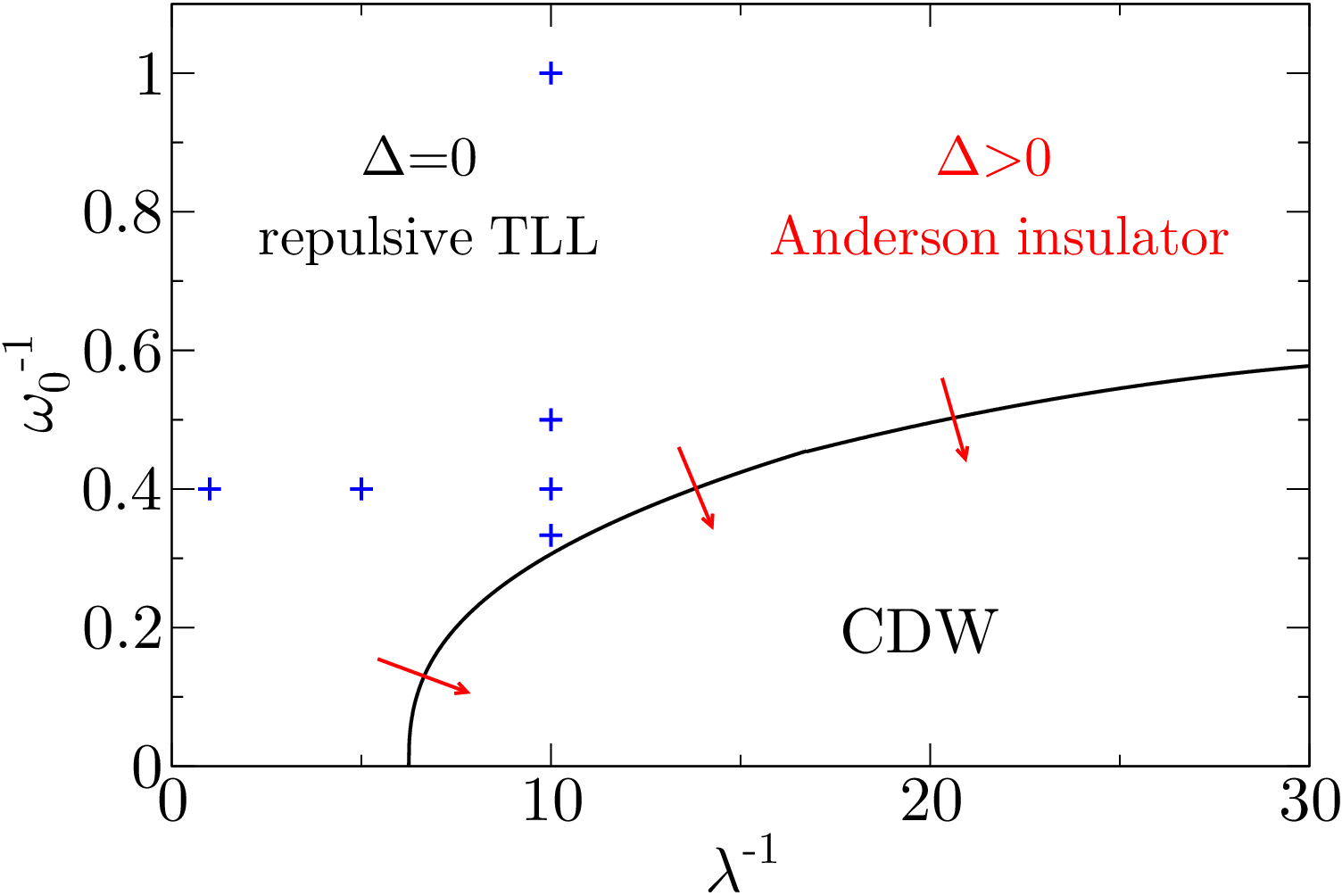}
\caption{(Color online) DMRG metal-insulator phase boundary 
for the 1D half-filled Edwards model without disorder (solid line). 
CDW order is suppressed if the background fluctuations dominate 
[$\omega_0 < 1$] or if the 
system's ability for relaxation is high [$\lambda >\lambda_c(\omega_0)$].
The blue crosses denote the parameter sets considered in this paper
for the disordered Edwards model.
}
\label{TLL-CDW-PD}
\end{figure}

The Edwards model~\cite{Ed06} represents a very general two-channel 
fermion-boson 
Hamiltonian, describing quantum transport in a background medium. 
Its fermion-boson interaction part       
\begin{equation}
 H_{fb}= -t_b\sum_{\langle i, j \rangle} f_j^{\dagger}f_{i}^{}
  (b_i^{\dagger}+b_j^{})
\label{fb-transport}
\end{equation} 
mimics the correlations/fluctuations inherent to a spinful fermion 
many-particle system by a boson-affected transfer of spinless 
charge carriers. In Eq.~(\ref{fb-transport}), a fermion $f_i^{(\dagger)}$ 
creates (or absorbs) a local boson $b_i^{(\dagger)}$ every time it hops to a 
nearest neighbor (NN) site $j$. Thereby it creates a local excitation in the 
background with energy $\omega_0$: $H_b=\omega_0\sum_i b_i^{\dagger}b_i^{}$. 
Because of quantum fluctuations the background distortions should be 
able to relax with a certain rate $\lambda$. The entire 
Edwards Hamiltonian then reads
\begin{equation}
 H_{E}= H_{fb} - \lambda\sum_i(b_i^{\dagger}+b_i^{}) + H_b\,.
\label{E-model1}
\end{equation} 
A unitary transformation $b_i\mapsto b_i+\lambda/\omega_0$ eliminates
the boson relaxation term in favor of a second fermion
hopping channel: 
\begin{equation}
 H_{E}= H_{fb} -t_f\sum_{\langle i, j \rangle} f_j^{\dagger}f_{i}^{} + H_b.
\end{equation} 
We like to emphasize that (i) this free-fermion transfer takes place
on a strongly reduced energy scale $t_{f}=2\lambda t_{\rm b}/\omega_0$
however, and (ii) coherent propagation of a fermion is possible 
even in the limit $\lambda=t_f=0$ by means of a six-step vacuum-restoring 
hopping process,~\cite{AEF07} acting as a direct next NN transfer 
``$f_{i+2}^{\dagger} f_i^{}$.'' The Edwards model reveals a surprisingly rich
physics. Depending on the relative strengths $t_f/t_b$ of the two transport 
mechanisms and the rate of bosonic fluctuations $t_b/\omega_0$ it reproduces
Holstein and $t$-$J$ model-like lattice- and spin-polaron transport, 
respectively, in the single-particle sector.~\cite{AEF07,BF10}      
For the half-filled band case, a metal-insulator quantum-phase transition 
from a repulsive Tomonaga-Luttinger-liquid (TLL) to a CDW has been 
reported,~\cite{WFAE08,EHF09} see Fig.~\ref{TLL-CDW-PD}.  
Note that the CDW is a few-boson state that typifies rather a correlated 
(Mott-Hubbard-type) insulator than a Peierls state with 
many bosons (phonons) involved.~\cite{WFAE08,EHF09} Since in the limit
$\omega_0 \gg 1 \gg \lambda$ (here, and in what follows $t_b$ is taken as
the unit of energy) background fluctuations are energetically costly, 
charge transport is hindered and an effective Hamiltonian with NN 
fermion repulsion results. To leading order, in a 
reduced (zero-boson) Hilbert space, we get   
\begin{equation}
H_{tV}= -t_f\sum_{\langle i, j \rangle} f_j^{\dagger}f_{i}^{}  + V \sum_i n_i^{f} n_{i+1}^{f}
\end{equation} 
with $V=t_{b}^2/\omega_0$. This so-called $t$-$V$ model can be mapped 
onto the exactly solvable $XXZ$ model, which exhibits a TLL-CDW 
quantum phase transition at $V/t_f = 2$, i.e., at $\lambda_{c}^{-1}=4$. 
This value is smaller than those obtained for the Edwards model 
in the limit $\omega_0^{-1}\ll 1$, where $\lambda_{c}^{-1} \simeq 6.3$ 
(see Fig.~\ref{TLL-CDW-PD} and Ref.~\onlinecite{EHF09}), because already 
three-site and effective next-NN hopping terms 
were neglected in the derivation of the $tV$-model.

We now employ the DMRG technique,~\cite{Wh92} which can be easily 
generalized to treat systems including bosons,~\cite{JW98b} in order to obtain 
unbiased results for the full AEM,  
\begin{equation}
 H_{AE}= \Delta \sum_i \varepsilon_i n_i^{f} + H_{E}\,,  
\label{AE-model}
\end{equation} 
and the related A$tV$M, $H_{AtV}=\Delta \sum_i \varepsilon_i n_i^{f} + H_{tV}$,  
where disorder of strength $\Delta$ is induced by independently distributed 
random on-site potentials $\varepsilon_i$, drawn from the box distribution 
$P(\varepsilon_i)=\theta(1/2-|\varepsilon_i|)$. Within the pseudosite 
approach, a boson is mapped to $n_{b}$ pseudosites.~\cite{JW98b,JF07}  
In the numerical study of the AEM we take into account 
up to $n_{b}=4$ pseudosites and determine $n_{b}$ by the requirement 
that local boson density of the last pseudosite is less 
than $10^{-7}$ for all $i$. Furthermore we keep up to $m=1200$ density-matrix 
eigenstates in the renormalization steps to ensure that the discarded weight 
is smaller than $10^{-8}$. The calculations are performed for finite 
systems with lengths $L=16$ to 128 and open boundary conditions (OBC).  
For the simpler effective A$tV$M we reach 
$L=192$ with OBC. Here the use of $m=1000$ density-matrix 
eigenstates makes the discarded weight negligible. To gain representative 
results for our disordered systems we proceed as follows. We first compute 
the physical quantity of interest at fixed $L$ for numerous samples 
$\{\varepsilon_i\}$, then set up an appropriate statistical average, 
and finally perform a careful finite-size scaling.      

\section{Finite-size scaling}
An important question is, of course, which physical quantity to use 
in the finite-size scaling of the Anderson transition. For this purpose 
the localization length $\xi$ seems to be promising, because it is sensitive to the 
nature, localized or extended, of the electron's eigenstate.~\cite{AALR79,KM93b}
So far $\xi$ has been determined from the phase sensitivity of the
ground-state energy.~\cite{SSSSE98,EB12} Quite recently,  Berkovits demonstrates 
that also the entanglement entropy can be used to extract the localization length.~\cite{Ber12} 
However, in both methods the system size $L$ should be always 
larger than $\xi$.

Advantageously, the localization length can be extracted by a finite-size 
scaling analysis of the charge density structure factor even for $L\ll\xi$, 
which works equally well for interacting systems,~\cite{NS10}
and therefore allows us to discuss the interplay between Anderson 
localization and CDW formation in a consistent manner. The charge 
structure is defined as
\begin{equation}
\tilde{C}(q)=\frac{1}{L}\sum_{i,r=1}^L[\langle n_i^{f} n_{i+r}^{f} \rangle - \langle n_i^{f} \rangle \langle n_{i+r}^{f} \rangle] e^{{\rm i}qr}.
\end{equation} 
Assuming an exponential decay of the equal-time density-density 
correlations in the Anderson insulating phase,~\cite{GS88} 
the structure factor scales with 
\begin{equation}
\tilde{C}(q)=-\frac{K_\rho^\ast}{2\pi^2} \frac{e^{-\frac{\pi^2L}{6\xi}}-1}{e^\frac{\pi^2}{6\xi}-1}q^2\,,
\label{fitting}
\end{equation}
where $q=2\pi/L\ll 1$.~\cite{NS10} Equation~(\ref{fitting}) contains two 
unknown parameters: the localization length $\xi$ and the 
disorder-modified TLL interaction coefficient $K_\rho^\ast$.  
Hence, if the charge structure factor is determined 
numerically, $\xi$ and $K_\rho^\ast$ can be easily derived by fitting 
the numerical data with Eq.~(\ref{fitting}). For vanishing disorder 
$\Delta \to 0$, $\xi$ diverges and $K_\rho^\ast$ becomes the ordinary TLL 
parameter $K_\rho$. We are aware that a disordered 1D system is 
no longer a TLL and consequently the TLL parameter is ill-defined in 
the strict sense. Nevertheless, if the localization length significantly 
exceeds the lattice constant, the short-range correlation functions 
should still show a power-law decay. Therefore we might gain 
some valuable information about the local motion of fermions 
from $K_\rho^\ast$. 

\section{DMRG results}
\begin{figure}[t]
\centering
\includegraphics[clip,width=.95\columnwidth]{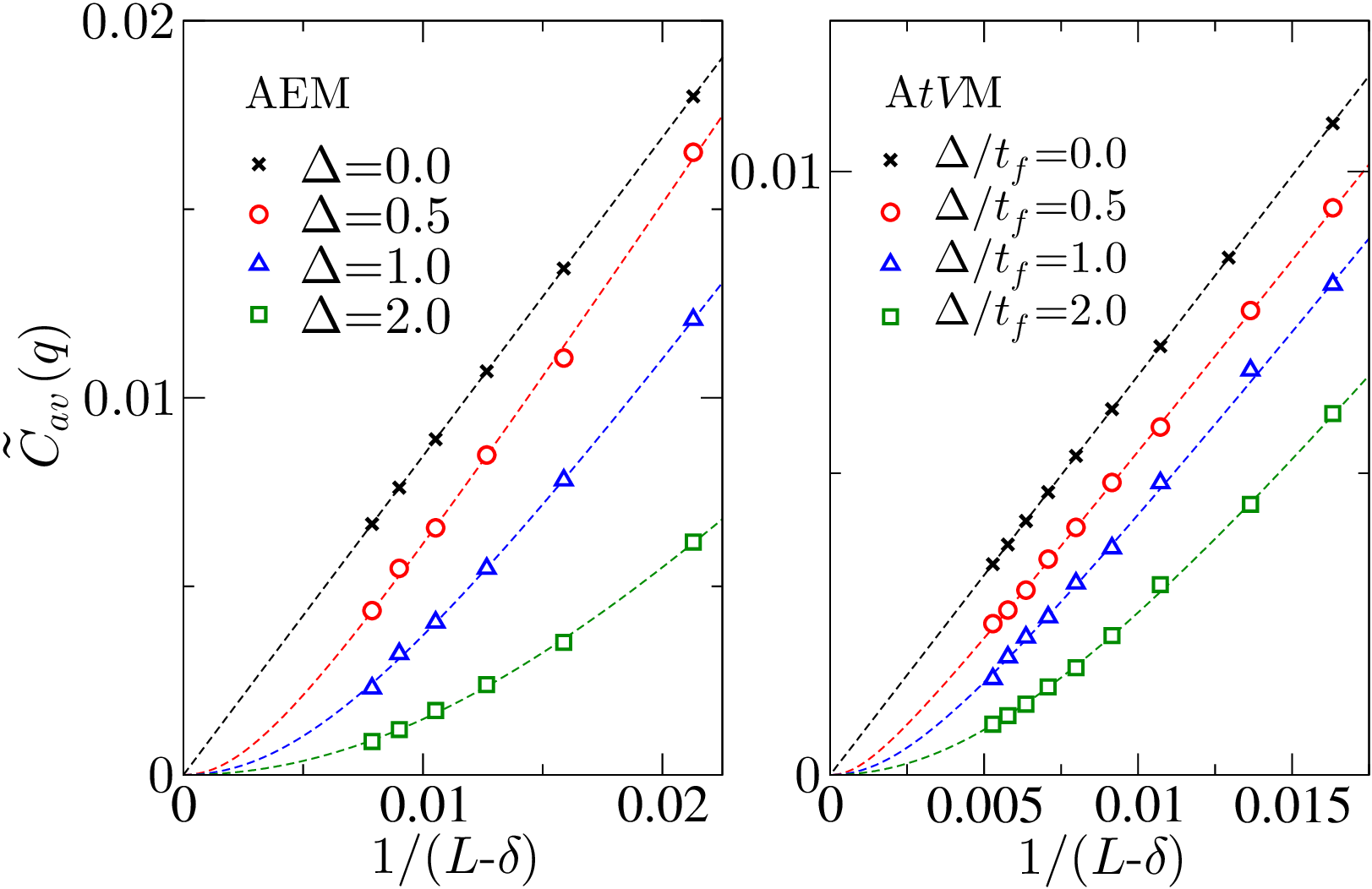}
\caption{(Color online) 
Charge structure factor of
the AEM with $\lambda=0.1$ and $\omega_0=2$ (left panel) and 
the A$tV$M with $V/t_f=1.5$ (right panel) sampled over 
300 and 500 disorder realizations, respectively. Dashed lines 
give the finite-size scaling of $\tilde{C}_{av}(q)$
according to~\eqref{fitting}.     
}
\label{CSF-FSS}
\end{figure}

Figure~\ref{CSF-FSS} demonstrates that the finite-size scaling of 
the averaged charge structure factor $\tilde{C}_{av}(q)$
by  means of~\eqref{fitting} works best and equally well for the 1D 
AEM and A$tV$M (this applies to all parameter values discussed below). 
To accommodate the missing correlations
owing to the OBC, we have plotted $\tilde{C}_{av}(q)$ as a function of 
$1/(L-\delta)$ instead of $1/L$ (this way of plotting the data is nonessential
but gives a quantitative refinement of the fit).
The parameter $\delta$ is adjusted to reproduce 
$K_\rho^\ast=K_\rho$ and $\xi =\infty$ at $\Delta=0$. We note the general 
tendency that the charge correlations arising at finite $L$ will be suppressed
as $\Delta$ becomes larger.    

\begin{figure}[t]
\centering
\includegraphics[clip,width=.95\columnwidth]{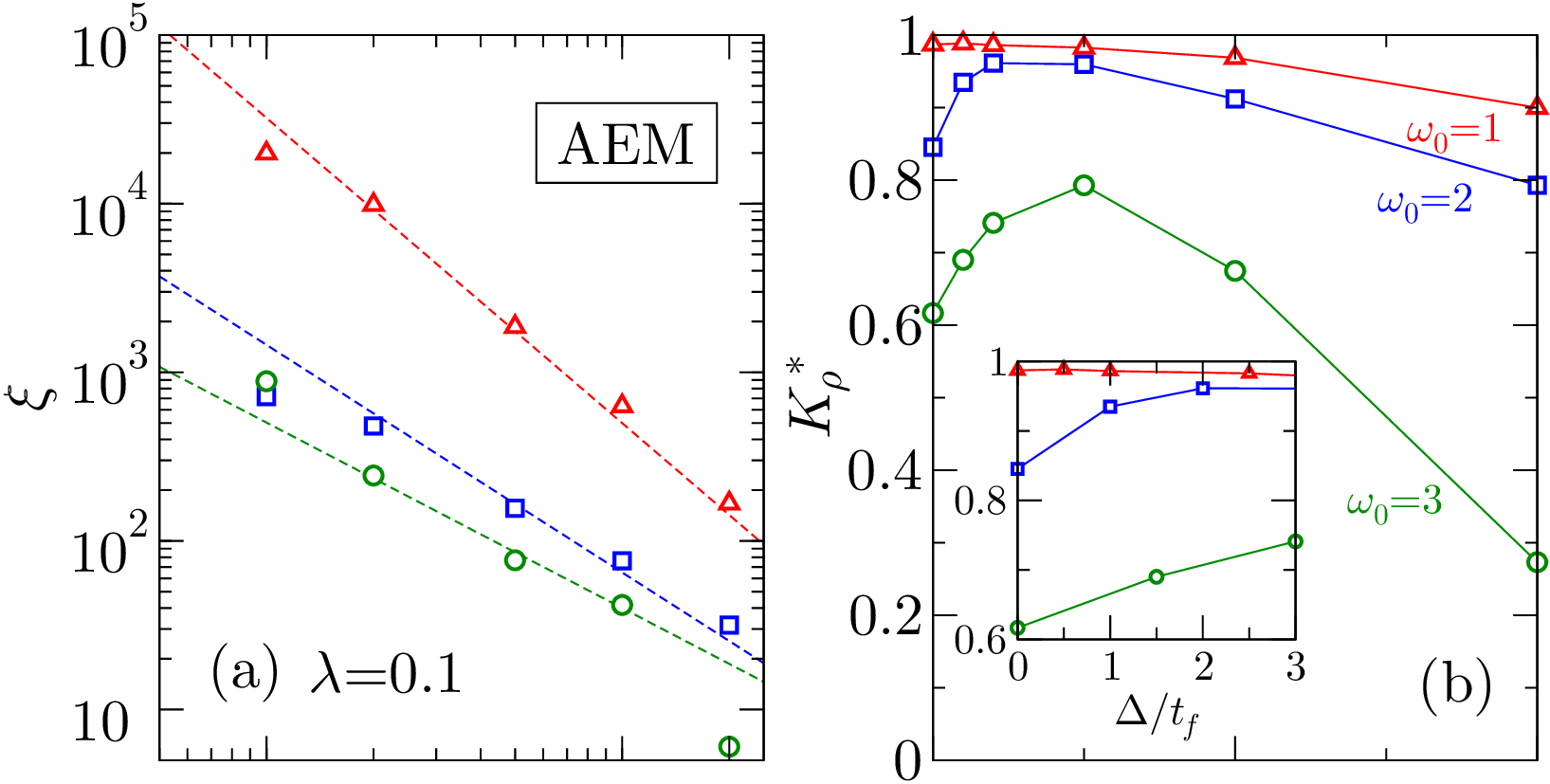}
\includegraphics[clip,width=.95\columnwidth]{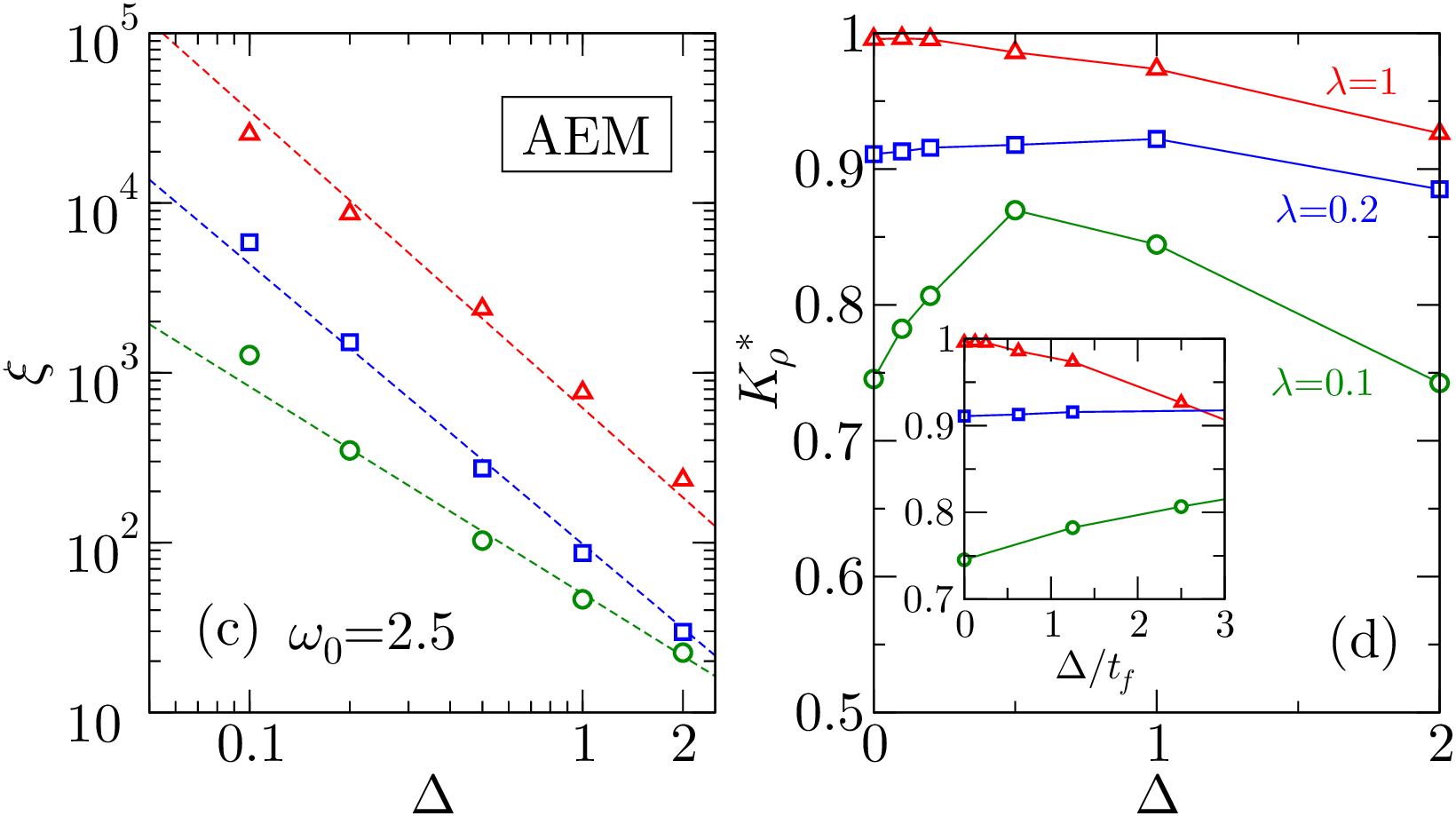}
\includegraphics[clip,width=.95\columnwidth]{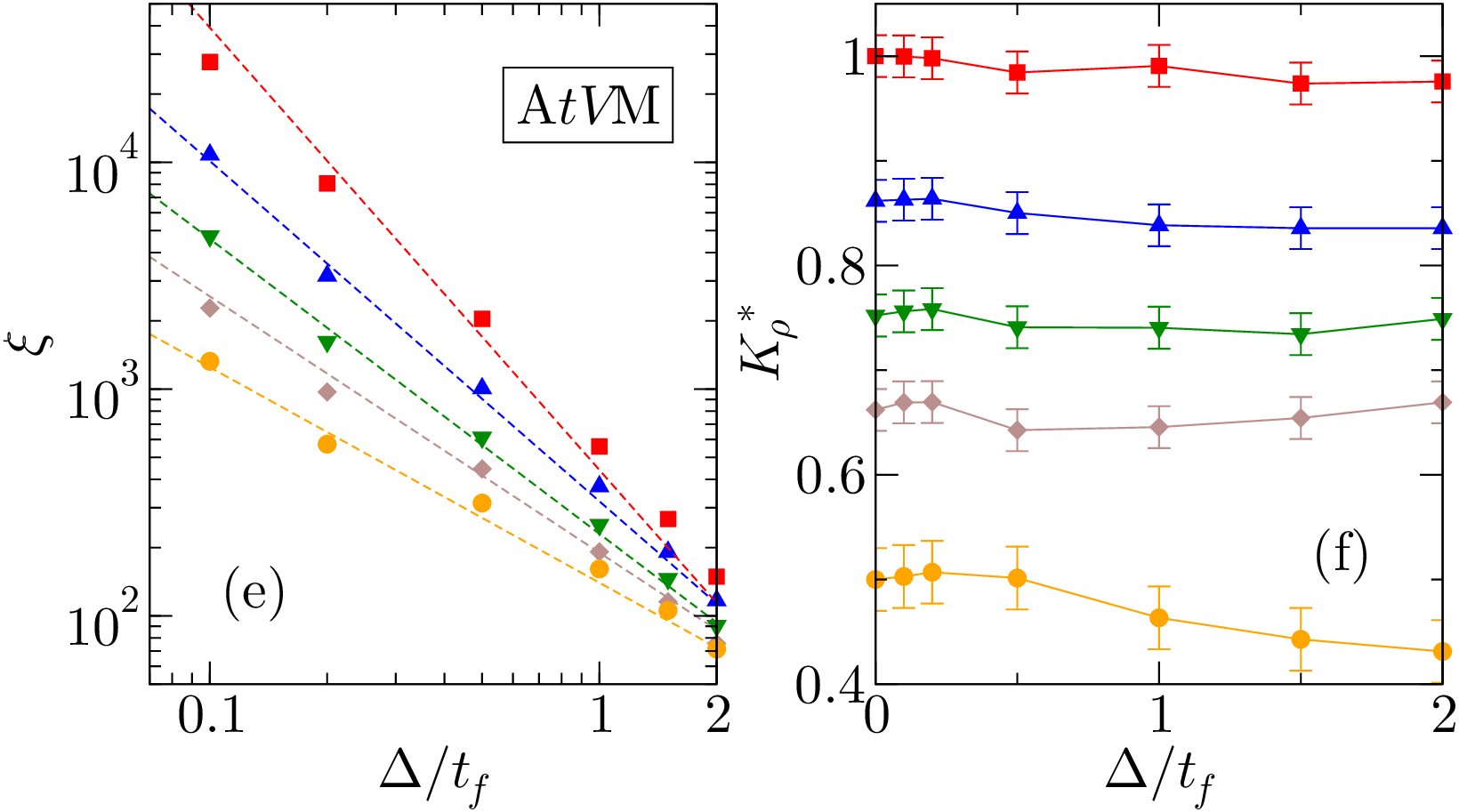}
\caption{(Color online) Left panels: Log-log plot of the localization 
length versus disorder strength for the AEM at fixed 
$\lambda=0.1$ (a), fixed $\omega_0=2.5$ (c), and for 
the A$tV$M (e). 
Dashed lines are fits to Eq.~\eqref{xi_general}
with:  (a) $\xi_0=620$, 98, 50 and $\gamma=1.75$, 1.65, 1.22 for
$\lambda=1$, 0.2, 0.1; (c) $\xi_0=500$, 65, 40 and 
$\gamma=1.81$, 1.35, 1.1 for $\omega_0=1$, 2, 3;
(e) $\xi_0=440$, 350, 230, 190, 150 and 
$\gamma=1.95$, 1.5, 1.3, 1.1, 0.95 for $V/t_f=0$, 0.5, 1. 1.5, 2 
(from top to bottom), respectively.  
Right panels: Corresponding results for
the modified TLL parameter $K_\rho^\ast$ in the AEM [(b), (d)], and
A$tV$M (f).}
\label{XI-TLL}
\end{figure}

In a next step, we extract the localization length $\xi$ and the modified 
TLL parameter for the disordered Edwards and $t$-$V$ models. 
Figure~\ref{XI-TLL} shows the dependence of $\xi$ and $K_\rho^\ast$ 
on the disorder strength $\Delta$. First of all, we find a power-law
decay of $\xi$ with $1/\Delta$ in the whole ($\lambda$, $\omega_0$; $V$) 
parameter regime:
 \begin{equation}
\xi/\xi_0=\Delta^{-\gamma}
\label{xi_general}
\end{equation}
[note that  $\Delta$ is given in 
units of $t_b$ ($t_f$) in Fig.~\ref{XI-TLL} for the AEM (A$tV$M)]. The estimated values of the (bare) decay length $\xi_0$ and 
the exponent $\gamma$ are given in the caption of Fig.~\ref{XI-TLL} 
for characteristic model parameters. 
As expected, the localization length decreases
with increasing disorder strength. Stronger electronic correlations,
i.e, smaller $\lambda$ or larger $\omega_0$ (larger $V$) in the 
AEM (A$tV$M), also tend to reduce $\xi$. 
In any case, $\xi$ turns out to be finite as soon as $\Delta>0$,
indicating that the repulsive TLL, if realized for $\Delta=0$, 
makes way for an Anderson insulator. Thereby the localization length
becomes comparable to the lattice spacing at $\Delta=2$ in the 
AEM with $\lambda=0.1$, $\omega_0=2.5$, 
while it is still about $10^2$ for the A$tV$M
with $V/t_f=2$. 

The right-hand panels display striking differences in 
the $\Delta$-dependence of $K_\rho^\ast$ for 
the models under consideration. These can be attributed to the fact,
that the Edwards model contains two energy scales $\lambda$ and 
$\omega_0$ while the physics of the $t$-$V$ model is merely governed 
by the ratio $V/t_f=t_b/2\lambda$, i.e,. $\omega_0$ drops out.
Far away from the CDW instability, however, both models describe a weakly 
correlated TLL with $K_\rho\lesssim 1$ and $K_\rho^\ast$ slowly 
decays as the disorder $\Delta$ increases (see the red open triangles
in  Fig.~\ref{XI-TLL}; to make the comparison with the $t$-$V$ model 
data easy, we have shown $K_\rho^\ast$ versus $\Delta/t_f$ in the insets). 
If we move towards the CDW instability by decreasing $\lambda$ at fixed
$\omega_0>\omega_{0,c}$ or increase $\omega_0$ at $\lambda<\lambda_c$ 
(cf. Fig.~\ref{TLL-CDW-PD}) a non-monotonous behavior develops.
At small $\Delta$,  $K_\rho^\ast$ is significantly enhanced as 
the disorder increases. Obviously weak disorder destabilizes the 2$k_F$-CDW 
correlations locally, since disorder-induced second- (and higher-) 
order boson-assisted (inelastic) hopping processes are possible 
in the AEM, even for $\omega_0\gg1$. This in 
sharp contrast to the A$tV$M, 
where only elastic scattering takes place and the intersite Coulomb repulsion 
is hardly affected by $\Delta$. As a result, in the disordered 
$t$-$V$ model the CDW correlations 
will be stronger and more robust. Hence, for the A$tV$M, $K_\rho^\ast$ 
appears to be nearly independent from $\Delta$ for  
$0.5 \lesssim V/t_f \lesssim 2$. 
This also notably differs from the behavior found for the disordered Hubbard 
model,~\cite{NS10} where the umklapp scattering is effectively 
enhanced by the formation of Mott plateaus appearing due to  
disorder.~\cite{OYTM08} If $\Delta$ exceeds a certain 
value in the AEM, $K_\rho^\ast$ starts to decrease and finally the whole scaling
procedure breaks down when $\xi\gtrsim 1$ (see the point at $\Delta=2$
in the upper right panel with $K_\rho^\ast$ well below 0.5).    
In this regime the wave functions of the particles are strongly
localized and the TLL-behavior is as much suppressed as the CDW 
correlations. Let us point out that the enhancement of $K_\rho^*$ 
triggered by the bosonic degrees of freedom might serve as an 
explanation for the observed increasing charge velocity near a negatively
charged defect in the single-wall carbon nanotubes,~\cite{MAPHQTTCRNJ08}
since the TLL parameter $K_\rho$ is proportional to the charge velocity.

\begin{figure}[t]
\centering
\includegraphics[clip,width=.95\columnwidth]{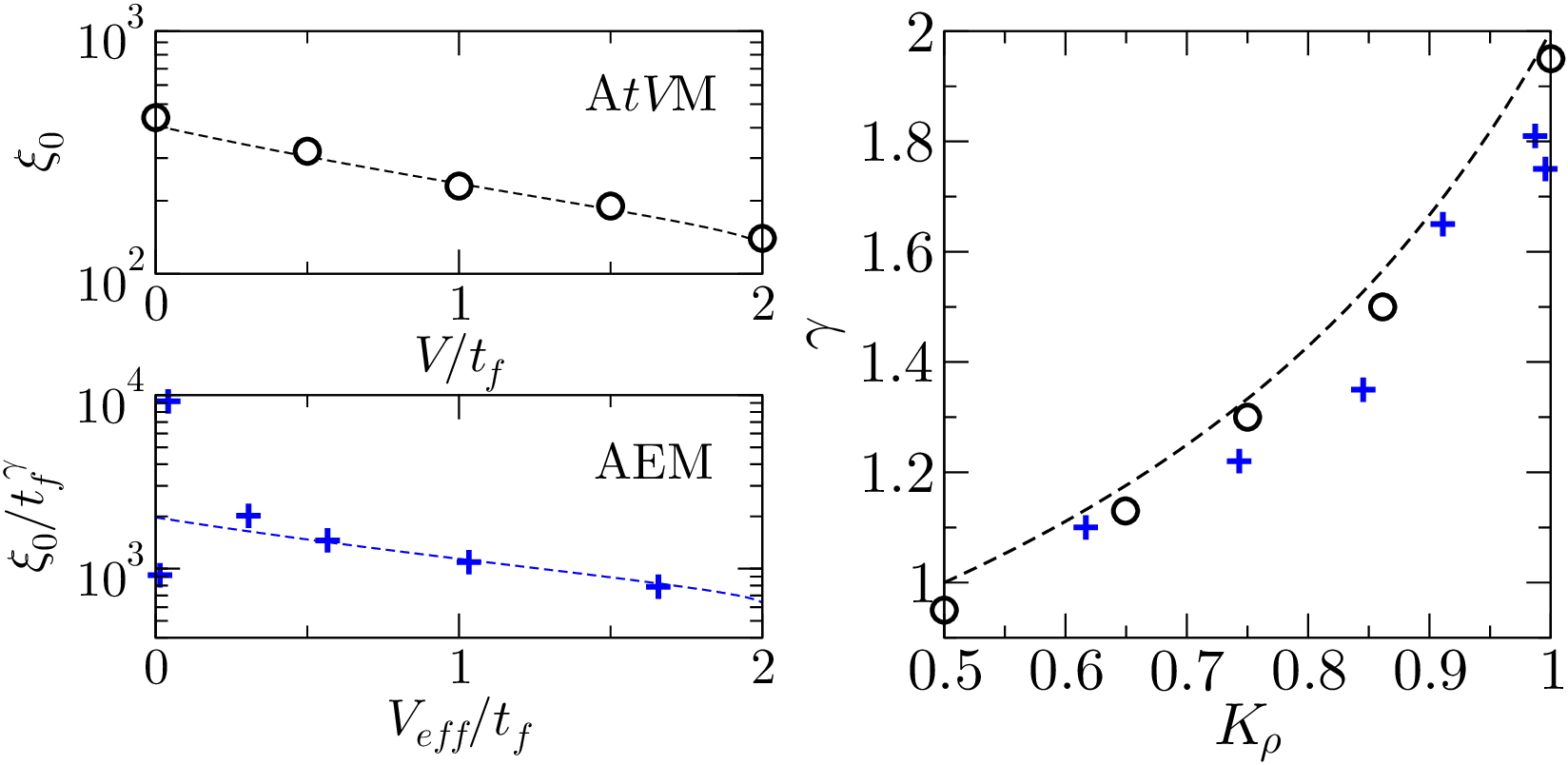}
\caption{(Color online) Left panels: Decay length $\xi_0$ 
as a function of the (effective) Coulomb repulsion $V_{(eff)}$
for the A$tV$M and AEM. Right panel: Corresponding 
$\gamma$-exponent {\it vs.} $K_\rho$, in comparison to the 
scaling relation~\eqref{sc_rel} (dashed line).
For further explanation see text. 
}
\label{XI0-GAMMA}
\end{figure}

We now focus on the localization behavior at large distances  
[${\cal O}(\xi\gg1)$], and therefore make an attempt to analyze 
the decay length $\xi_0$ and the exponent $\gamma$, 
for both the AEM and A$tV$M, in terms of the interaction exponent 
$K_\rho$ and the charge susceptibility $\chi_c$ of standard 
TLL theory.~\cite{GS88} We expect that $\xi_0$ is strongly 
influenced by the strength of the charge fluctuations quantified by 
$\chi_c$, which is given---for the $t$-$V$ model---as 
\begin{equation}
\chi_{c}=\frac{2K_\rho}{\pi v_{c}}=\frac{4}{\pi
 \sqrt{1-(\frac{V}{2t_f})^2}}
\bigg[\frac{\pi}{\arccos(-\frac{V}{2t_f})}-1\bigg]
\end{equation} 
with charge velocity $v_{c}$. Figure~\ref{XI0-GAMMA} shows that
$\xi_0$ nicely scales with $V/t_f$, i.e., $\xi_0\propto \chi_c$ in fact
(see upper panel). The same holds---perhaps surprisingly---for the 
AEM when $\xi_0/t_f^\gamma$ 
is plotted versus an effective intersite repulsion 
estimated from $V_{eff}/t_f=-2 \cos (\pi/2K_\rho)$ (see lower panel). 
Moreover, for the disordered $t$-$V$ model, the exponent 
$\gamma$ is connected to $K_\rho$ of the (spinless fermion) 
TLL system without disorder via the renormalization equation: 
$d(\Delta^2)/dl=(3-2K_\rho)\Delta^2$ (with scale quantity $l$). 
This causes the scaling relation~\cite{GS88,WE87,HQB87}
\begin{equation}
\gamma=2/(3-2K_\rho)\,.
\label{sc_rel}
\end{equation}
The right panel of Fig.~\ref{XI0-GAMMA} displays that $\gamma$ basically 
depends on $K_\rho$ as predicted by Eq.~\eqref{sc_rel}. This means that
the long-range localization properties of the AEM can be understood 
in the framework of A$tV$M with an effective intersite interaction
induced by the bosonic degrees of freedom. Since the (effective)
Coulomb repulsion tends to result in a lesser $K_\rho$, $\gamma$ decreases
with increasing $V_{(eff)}$ (cf. Fig.~\ref{XI0-GAMMA}). 
In this way the 2$k_F$-CDW fluctuations triggered by $V$ tend to weaken Anderson localization. 
While $\gamma=2$ in the free-fermion limit ($V,\,1/\lambda\to 0$), 
it scales to unity approaching the CDW transition point, located, e.g., 
at $\lambda_c\simeq 0.07$ for $\omega_0=2.5$ respectively 
at $\omega_{0,c}\simeq 3.1$ for $\lambda=0.1$.   

\begin{figure}[t]
\centering
\includegraphics[clip,width=.9\columnwidth]{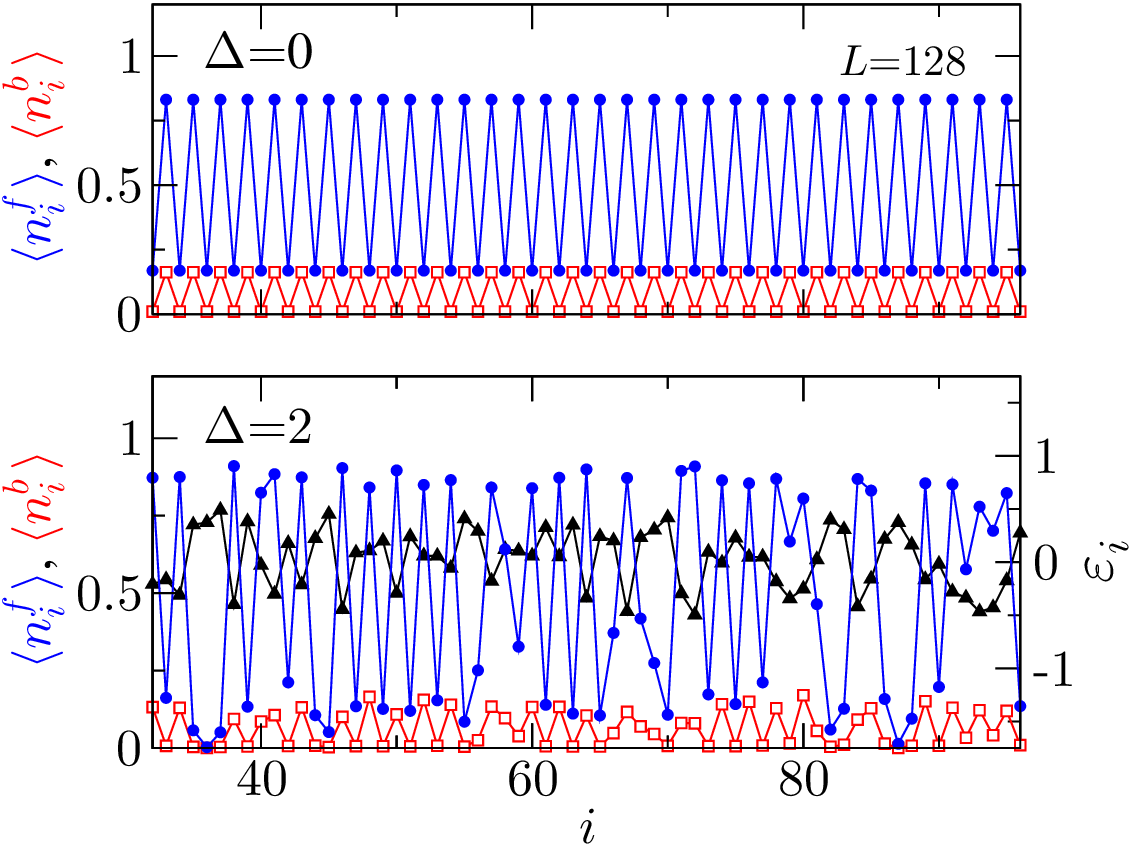}
\caption{(Color online) Local densities of fermions 
$\langle f_i^\dagger f_i^{}\rangle$ (blue circles) and bosons 
$\langle b_i^\dagger b_i^{}\rangle$ (red open squares)
in the central part of an open AEM chain without
($\Delta=0$) and with ($\Delta=2$) disorder.
Results are given for a single realization 
$\{\varepsilon_i\}$ (black triangles). Model parameters are $\lambda=0.01$,
$\omega_0=2.5$, and $L=128$ (OBC).   
}
\label{NF-NB}
\end{figure}

The question of how disorder affects the insulating CDW state could not 
be addressed by the above TLL-based scaling. In particular,
we cannot assess by our numerical approach whether the CDW phase survives
{\it weak} disorder (as experimentally observed for disordered 
Peierls-Mott insulators).~\cite{WOHPKL12} If the Imry-Ma argument 
for disordered (low-D) interacting systems~\cite{IM75}
holds, CDW long-range order should be destroyed by any finite $\Delta$. 
Figure~\ref{NF-NB}, showing the spatial variation of the local fermion/boson 
densities for a specific but typical disorder realization 
(note that any real experiment is performed on a particular sample), 
illustrates the situation deep inside the (former) CDW phase ($\lambda^{-1}=100$,
$\omega_0^{-1}=0.4$; cf. Fig.~\ref{TLL-CDW-PD}). One realizes
that long-range charge order ceases to exist but short-range CDW 
correlations locally persist whenever neighboring on-site potentials
do not differ much (see, e.g., the region $i=45\ldots 55$ in the lower 
panel of Fig.~\ref{NF-NB}).   
 
\section{Conclusions}
To summarize, using an unbiased numerical DMRG approach, 
we investigated the interplay of disorder 
and interaction effects including bosonic degrees of freedom 
in the framework of the 1D spinless fermion Anderson-Edwards model. 
Although the TLL phase disappears owing to the disorder,
the localization properties of the Anderson insulator state 
can be understood in terms of scaling relations containing 
the charge susceptibility and the Luttinger liquid parameter 
of the metallic phase without disorder only, as in the case of
the spinless fermion Anderson-$t$-$V$ model. However,
the Anderson-Edwards model reveals a more complex interrelation  
between disorder and CDW correlations because additional 
scattering channels, involving bosonic excitation and 
annihilation processes, appear. This offers a promising route for adapting  
the description of low-dimensional transport in many disordered materials.  
Disorder also affects the CDW state in that true long-range order vanishes 
although local CDW correlations 
survive. 

\acknowledgments
This work was supported by DFG through SFB 652.

\bibliography{ref} 
\bibliographystyle{apsrev}

\end{document}